\documentclass{article}
\usepackage[cp866]{inputenc}
\begin{document}

\centerline {{\Large\bf Specific features of differential equations }}
\centerline {{\Large\bf of mathematical physics}}
\centerline {\it L.~I. Petrova}

\renewcommand{\abstractname}{Abstract}
\begin{abstract}
Three types of equations of mathematical physics, namely, the equations, 
which describe any physical processes, the equations of mechanics
and physics of continuous media, and field-theory equations are studied
in this paper.

In the first and second case the investigation is reduced to the analysis
of the nonidentical relations of the skew-symmetric differential forms that
are obtained from differential equations. It is shown that the integrability
of equations and the properties of their solutions depend on the realization of
the conditions of degenerate transformations under which the identical 
relations are obtained from the nonidentical relation.

The field-theory equations, in contrast to the equations of first two types,
are the relations made up by skew-symmetric differential forms or their
analogs (differential or integral ones). This is due to the fact that the
field-theory equations have to describe physical structures (to which
closed exterior forms correspond) rather than physical quantities. 
The equations that correspond to field theories are obtained from 
the equations that describe the conservation laws (of energy, linear 
momentum, angular momentum, and mass) of material systems (of continuous 
media). This disclose a connection between field theories and the 
equations for material systems (and points to that material media 
generate physical fields).

\end{abstract}

\subsection*{1. Specific features of equations descriptive of physical
processes}

Specific features of differential equations descriptive of physical
processes and the types of their solutions can be demonstrated by the
example of first-order partial differential equation using the
properties of skew-symmetric differential forms.

{\footnotesize [The method of investigating differential equations
using skew-symmetric differential forms was developed
by Cartan [1] in his analysis of the integrability of differential
equations. Here we present this analysis to demonstrate specific features
of differential equations and properties of solutions to these equations.]}

Let
$$ F(x^i,\,u,\,p_i)=0,\quad p_i\,=\,\partial u/\partial x^i \eqno(1)$$
be a first-order partial differential equation.

Let us consider the functional relation
$$ du\,=\,\theta\eqno(2)$$
where $\theta\,=\,p_i\,dx^i$ (the summation over repeated indices is
implied). Here $\theta\,=\,p_i\,dx^i$ is a differential form of the
first degree.

The specific feature of functional relation (2) is that in the general
case, when differential equation (1) describes any
physical processes, this relation turns out to be nonidentical one.

The left-hand side of this relation involves a differential, and
the right-hand side includes the differential form
$\theta\,=\,p_i\,dx^i$. For this relation be identical, the
differential form $\theta\,=\,p_i\,dx^i$ must also  be a differential
(like the left-hand side of relation (2)), that is, it has to be a
closed exterior differential form. To do this, it requires the commutator
$K_{ij}=\partial p_j/\partial x^i-\partial p_i/\partial x^j$ of the
differential form $\theta $ has to vanish.

In the general case
from equation (1) it does not follow (explicitly)
that the derivatives $p_i\,=\,\partial u/\partial x^i $, which obey
to the equation (and given boundary or initial conditions of the
problem), make up a differential. For equations descriptive of
any processes (without any supplementary conditions),
the commutator $K_{ij}$ of the differential form $\theta $
is not equal to zero. The form $\theta\,=\,p_i\,dx^i$ turns out to be
unclosed and is not a differential like the left-hand side of relation
(2). Functional relation (2) appears to be nonidentical.

The nonidentity of functional relation (2) points to a fact
that without additional conditions the derivatives of original
equation do not make up a differential. This means that the
corresponding solution $u$ of the differential equation will not be
a function of only variables $x^i$. The solution will depend on
the commutator of the form $\theta $, that is, it will be a functional.

To obtain a solution that is a function (i.e., the derivatives of
this solution make up a differential), it is necessary to add the
closure condition for the form $\theta\,=\,p_idx^i$ and for
relevant dual form (in the present case the functional $F$
plays a role of a form dual to $\theta $) [1]:
$$\cases {dF(x^i,\,u,\,p_i)\,=\,0\cr
d(p_i\,dx^i)\,=\,0\cr}\eqno(3)$$

If we expand the differentials, we get a set of homogeneous equations
with respect to $dx^i$ and $dp_i$ (in the $2n$-dimensional tangent
space):
$$\cases {\displaystyle \left ({{\partial F}\over {\partial x^i}}\,+\,
{{\partial F}\over {\partial u}}\,p_i\right )\,dx^i\,+\,
{{\partial F}\over {\partial p_i}}\,dp_i \,=\,0\cr
dp_i\,dx^i\,-\,dx^i\,dp_i\,=\,0\cr} \eqno(4)$$

It is well-known that {\it vanishing the determinant} composed of coefficients
at $dx^i$, $dp_i$ is a solvability condition of the system of homogeneous
differential equations. This leads to relations:
$$
{{dx^i}\over {\partial F/\partial p_i}}\,=\,{{-dp_i}\over {\partial F/\partial x^i+p_i\partial F/\partial u}} \eqno (5)
$$

Relations (5) specify the integrating direction. namely, a pseudostructure,
on which the form $\theta \,=\,p_i\,dx^i$ turns out to be closed one,
i.e. it becomes a differential, and from relation (2) the identical relation
is produced.  One the pseudostructure, which is  defined by relation(5),
the derivatives of differential equation (1) constitute
a differential $\delta u\,=\,p_idx^i\,=\,du$ (on the pseudostructure), and
the means that the solution of equation (1) becomes a function.

Solutions, namely, functions on the pseudostructures
formed by the integrating directions, are the so-called generalized
solutions.

{\footnotesize [If we find the characteristics of equation (1), it appears that
relations (5) are characteristic relations [2].
That is, the characteristics are examples of the pseudostructures on
which the derivatives of the differential equation made up closed forms
and the solutions prove to be functions (generalized solutions).]}

If the requirements of closure of skew-symmetric form made up by
the derivatives of differential equation and relevant dual form
are not fulfilled, that is, the derivatives do not form a
differential, the solution corresponding to such derivatives will
depend on the differential form commutator formatted by
derivatives. That means that the solution is a functional rather
than a function.

The first-order partial differential equation has been analyzed,
and the functional relation with the form of the first degree
has been considered.

Similar functional properties have the solutions to all differential
equations describing physical processes. And, if the order of the
differential equation is $k$, the functional relation with the
$k$-degree form corresponds to this equation.

\bigskip
Thus one can see that the differential equations describing any
physical fields can have solutions of two types, namely,
generalized solutions which depend on variables only, and the solutions
which are functionals since they depend on the commutator made up
by mixed derivatives. A specific feature of generalized solutions
consists in the fact that they can be realized only under {\it
degenerate transformations}. The relations (5) corresponding to
generalized solutions had been obtained the condition of {\it
vanishing the determinant} composed of coefficients at $dx^i$,
$dp_i$ in the set of equations (4). This is a condition of {\it
degenerate} transformation. (They are connected with symmetries of
commutators of skew-symmetric forms.)

{\footnotesize [It is clear that the degenerate transformation is a transition
from tangent space to cotangent space (the Legendre transformations).
The coordinates in relations (5) are not identical to the independent
coordinates of the initial space on which equation (1) is defined.]}

Since generalized solutions are possible only under realization of
the conditions of degenerate transforms, they are discrete
solutions (defined only on pseudostructures) and have
discontinuities  in the direction normal to pseudostructures.

The solutions being functionals disclose the another peculiarity
of the solutions of differential equations, namely, their
instability. The dependence of the solution on the commutator may
lead to instability. The instability develops when the
integrability conditions are not realized and exact (generalized)
solutions are not formatted. (Thus, the solutions to the equations
of the elliptic type may be unstable.)

{\footnotesize [One can see that the qualitative theory of
differential equations that solves the problem of unstable
solutions and integrability bases on the properties nonidentical
functional relation.]}

\subsection*{2. Peculiarities of differential equations of mechanics and physics
of material media }

In analysis of partial differential equations the conjugacy of
derivatives in different directions was studied (using the
nonidentical functional relation). Under description of physical
processes in material (continuous) media one obtains not one
differential equation but a set of differential equations. And in
this case it is necessary to investigate the conjugacy of not only
derivatives in different directions  but also the conjugacy
(consistency) of the equations of this set. In this case from
this set of equations one also obtains  nonidentical relation that
allows to study the conjugacy of equations and features of their
solutions.

{\footnotesize [The material (continuous) medium - material system
- is a variety (infinite) of elements  that have internal
structure and interact among themselves. Thermodynamical,
gasodynamical and cosmologic system, systems of elementary
particles and others are examples of material system. (Physical
vacuum can be considered as an analog of such material system.)
Electrons, protons, neutrons, atoms, fluid particles and so on are
examples of elements of material system.]}

Equations of mechanics and physics of continuous media are
equations  that describe the conservation laws for energy, linear
momentum, angular momentum and mass. Such conservation laws can be
named as balance ones since they establish the balance  between
the variation of a physical quantity and corresponding external
action.

The equations of balance conservation laws are
differential (or integral) equations that describe a variation of functions
corresponding to physical quantities [3-5]. (The Navier-Stokes equations are
an example [5].)

(Mechanics and physics of continuous media treat the same equations.  However
an approach to solving these equations in mechanics and physics are different.
Below  it will be shown in what this difference manifest themselves.)

It appears that, even without a knowledge of the concrete form of these
equations, one can see specific features of these equations and their
solutions using skew-symmetric differential forms.

To do so it is necessary to study the conjugacy (consistency) of
these equations.

The functions for equations of material media sought are usually
functions which relate to such physical quantities like a particle
velocity (of elements), temperature or energy, pressure and
density. Since these functions relate to one material system, it
has to exist a connection between them. This connection is
described by the state-function. Below it will be shown that the
analysis of integrability and consistency of equations of balance
conservation laws for material media reduces to a study the
nonidentical relation for the state-function.

Let us analyze the equations that describe the balance conservation
laws for energy and linear momentum.

We introduce two frames of reference: the first is an inertial one
(this frame of reference is not connected with the material system), and
the second is an accompanying
one (this system is connected with the manifold built by
the trajectories of the material system elements).

The energy equation
in the inertial frame of reference can be reduced to the form:
$$
\frac{D\psi}{Dt}=A_1
$$
where $D/Dt$ is the total derivative with respect to time, $\psi $ is
the functional of the state that specifies the material system, $A_1$ is
the quantity that
depends on specific features of the system and on external energy
actions onto the system.
{\footnotesize \{The action functional, entropy, wave function
can be regarded as examples of the functional $\psi $. Thus, the
equation for energy presented in terms of the action functional $S$ has
a similar form:
$DS/Dt\,=\,L$, where $\psi \,=\,S$, $A_1\,=\,L$ is the Lagrange function.
In mechanics of continuous media the equation for
energy of an ideal gas can be presented in the form [5]: $Ds/Dt\,=\,0$, where
$s$ is entropy.\}}

In the accompanying frame of reference the total derivative with respect to
time is transformed into the derivative along the trajectory. Equation
of energy is now written in the form
$$
{{\partial \psi }\over {\partial \xi ^1}}\,=\,A_1 \eqno(6)
$$
Here  $\xi^1$ is the coordinate along the trajectory.

In a similar manner, in the accompanying reference system the
equation for linear momentum appears to be reduced to the equation of
the form
$$
{{\partial \psi}\over {\partial \xi^{\nu }}}\,=\,A_{\nu },\quad \nu \,=\,2,\,...\eqno(7)
$$
where $\xi ^{\nu }$ are the coordinates in the direction normal to the
trajectory, $A_{\nu }$ are the quantities that depend on the specific
features of material system and on external force actions.

Eqs. (6) and (7) can be convoluted into the relation
$$
d\psi\,=\,A_{\mu }\,d\xi ^{\mu },\quad (\mu\,=\,1,\,\nu )\eqno(8)
$$
where $d\psi $ is the differential
expression $d\psi\,=\,(\partial \psi /\partial \xi ^{\mu })d\xi ^{\mu }$.

Relation (8) can be written as
$$
d\psi \,=\,\omega \eqno(9)
$$
here $\omega \,=\,A_{\mu }\,d\xi ^{\mu }$ is the skew-symmetrical differential
form of the first degree.

Relation (9) has been obtained from the equation of the balance conservation
laws for energy and linear momentum. In this relation the form $\omega $
is that of the first degree. If the equations of the balance conservation
laws for angular momentum be added to the equations for energy and linear
momentum, this form  will be a form of the
second degree. And in combination with the equation of the balance
conservation law for mass this form will be a form of degree 3.
In general case the evolutionary relation can be written as
$$
d\psi \,=\,\omega^p \eqno(10)
$$
where the form degree  $p$ takes the values $p\,=\,0,1,2,3$.
{\footnotesize (The relation for $p\,=\,0$ is an analog to
that in the differential forms, and it was obtained from the
interaction of energy and time.)}

Since the balance conservation laws are evolutionary ones, the relations
obtained are also evolutionary relations, and the skew-symmetric forms
$\omega $ and $\omega^p $ are evolutionary ones.

Relations obtained from the equation of the balance conservation
laws, as well as functional relation (2), turn out to be nonidentical.

To justify this we shall analyze relation (9). This relation proves to be
nonidentical since the left-hand side of the relation is a differential,
which is a closed skew-symmetric form, but the right-hand side of the
relation involves the skew-symmetric differential form $\omega$, which
is unclosed form. The commutator made up by the derivatives of
coefficients $A_{\mu }$ the form $\omega $ itself is also nonzero, since
the coefficients $A_{\mu }$ are of different nature, that is, some
coefficients have been obtained from the energy equation and depend on
the energetic actions, whereas the others have been obtained from the
equation for linear momentum and depend on the force actions.

In a similar manner one can prove the nonidentity of relation (10).

Nonidentity of the evolutionary relation, as well as nonidentity on functional
relation (2), means that initial equations of balance conservation laws
are not conjugated, and hence they are not integrable. The solutions of
these equations can be functional or generalized ones. In this case
generalized solutions are obtained only under degenerated
transformations.

A type of solutions of the balance conservation law equations
is essential to mechanics and physics of continuous media. And
in physics and mechanics the interest is expressed in different types of
solutions. In physics the interest is expressed in only generalized 
solutions that are
invariant ones, and noninvariant solutions are ignored (even they
have a physical meaning). In contrast to this, in mechanics of
continuous media, where typically the equations are solved
numerically, one searches for solutions that are functionals. In
this case the question of searching for invariant solutions that
are realized only under additional conditions is not posed.

Such limited approach to solving the equations of material media has
some negative points.
The physical approach enables one to find possible invariant solutions,
however in this approach there is no way of telling in what instant of time
of evolutionary process one or another solution is obtained. This does not
also discloses the causality of phenomenon described by these solutions.
The approach exploited in mechanics of continuous media leads to
difficulties in explaining such phenomena as origination any discrete formations
(like a generation of waves or turbulent pulsations, birth of massless particles
and so on), to which the invariant solutions are assigned.
(The answer to the questions arisen in physics and mechanics while solving
the equations describing material media can be found by analysis of the
nonidentical evolutionary relation obtained from these equations.)

The evolutionary relation obtained from equations of balance
conservation laws for material systems (continuous media), in
contrast to functional relation (2), carries not only mathematical
but also large physical loading [6,7]. This is due to the fact that the 
evolutionary relation possesses the duality. On the one hand, this 
relation corresponds to material system, and on other, as it will be 
shown below, describes the mechanism of generating physical structures. 
This discloses the properties and peculiarities of the
field-theory equations and their connection with the equations of
balance conservation laws.

\bigskip

{\bf Physical significance of nonidentical evolutionary relation.}

The evolutionary relation describes the evolutionary process in material system
since this relation includes the state differential, which specifies
the material system state.
However, since this relation turns out to be not identical, from this
relation one cannot get the differential $d\psi $. The absence of
differential means that the system state is nonequilibrium.

The evolutionary relation possesses one more peculiarity,
namely, this relation is a selfvarying relation. (The evolutionary form
entering into this relation is defined on the deforming manifold
made up by trajectories of the material system elements. This means
that the evolutionary form basis varies. In turn,
this leads to variation of the evolutionary form, and the process
of intervariation of the evolutionary form and the basis is
repeated.)

Selfvariation of the nonidentical evolutionary relation points to the
fact that the nonequilibrium state of material system turns out
to be selfvarying. {\footnotesize (It is evident that this selfvariation proceeds under the action of
internal force whose quantity is described by the commutator of the
unclosed evolutionary form $\omega^p $.)}
State of material system changes but remains
nonequilibrium during this process.

Since the evolutionary form is unclosed, the evolutionary relation cannot be
identical. This means that the nonequilibrium state of material
system holds. But in this case it is possible a transition of material
system to a locally equilibrium state.

This follows from one more property of nonidentical
evolutionary relation. Under selfvariation of the evolutionary relation
it can be realized the conditions of degenerate transformation.
And under degenerate transformation from the nonidentical relation it
is obtained the identical relation.

From identical relation one can define the state differential pointing
to the equilibrium state of the system. However, such system state is
realized only locally due to the fact that the state differential
obtained is an interior one defined only on pseudostructure, that is
specified by the conditions of degenerate transformation. And yet
the total state of material system remains to be
nonequilibrium because the evolutionary relation, which describes
the material system state, remains nonidentical one.

The conditions of degenerate transformation are connected with symmetries 
caused by degrees of freedom of material system. These are symmetries of
the metric forms commutators of the manifold.  {\footnotesize \{To the
degenerate transformation it must correspond a vanishing of some
functional expressions, such as Jacobians, determinants, the Poisson
brackets, residues and others. Vanishing of these
functional expressions is the closure condition for dual form.
And it should be emphasize once more that the degenerate
transformation is realized as a transition from the accompanying
noninertial frame of reference to the locally inertial system. The
evolutionary form and nonidentical evolutionary relation are defined in
the noninertial frame of reference (deforming manifold). But the closed
exterior form obtained and the identical relation are obtained
with respect to the locally-inertial frame of reference
(pseudostructure)\}}.

Realization of the conditions of degenerate transformation is a vanishing
of the commutator of manifold metric form, that is, a vanishing of
the dual form commutator. And this leads to realization of pseudostructure
and formatting the closed inexact form, whose closure conditions have the form
$$d_\pi \omega^p=0,  d_\pi{}^*\omega^p=0 \eqno(12)$$

On the pseudostructure $\pi$ from evolutionary relation (10) it is 
obtained the relation
$$
d_\pi\psi=\omega_\pi^p\eqno(13)
$$
which proves to be an identical relation  since the closed inexact form
is a differential (interior on pseudostructure).

The realization of the conditions of degenerate transformation and
obtaining identical relation from nonidentical one has both
mathematical and physical meaning. Firstly, this points to the
fact that the solution of equations of balance conservation laws
proves to be a generalized one. And secondly, from this relation
one obtains the differential $d_\pi\psi $ and this points to the
availability of the state-function (potential) and that the state
of material system is in local equilibrium.

\bigskip
Relation (13) holds the duality. The left-hand side of relation
(13) includes the differential, which specifies material system
and whose availability points to the locally-equilibrium state of
material system. And the right-hand side includes a closed inexact
form, which is a characteristics of physical fields. The closure
conditions (12) for exterior inexact form correspond to the
conservation law, i.e. to a conservative on pseudostructure
quantity, and describe a differential-geometrical structure. These
are such structures (pseudostructures with conservative
quantities) that are physical structures formatting physical
fields[6].

The transition from nonidentical relation (10) obtained from
the balance conservation laws to identical
relation (13) means the following. Firstly, an emergency of the
closed (on pseudostructure) inexact exterior form (right-hand side
of relation (13)) points to an origination of the physical structure.
And, secondly, an existence of the state differential (left-hand side
of relation (13)) points to a transition of the material system from nonequilibrium state
to the locally-equilibrium state.

Thus one can see that the transition of material system from
nonequilibrium state to locally-equilibrium state is accompanied
by originating differential-geometrical structures, which are
physical structures.  Massless particles, charges,
structures made up by eikonal surfaces and wave fronts, and so on are
examples of physical structures.

The duality of identical relation also explains the duality of 
nonidentical evolutionary relation. On the one hand, evoltionary 
relation describes the evolutionary process in material systems, 
and on the other describes the process of generating physical fields.

Such duality, which establishes the connection between material
systems and physical fields, discloses one more peculiarity of
evolutionary processes in material media.

The emergency of physical structures in the evolutionary process
reveals in material system as an emergency of certain observable
formations, which develop spontaneously. Such formations and their
manifestations are fluctuations, turbulent pulsations, waves, vortices,
and others. It appears that structures of physical fields and the 
formations of material systems observed are a manifestation of the same 
phenomena. The light is an example of such a duality. The light 
manifests itself in the form of a massless particle (photon) and of 
a wave.

This duality also explains a distinction in studying the same phenomena
in material systems and physical fields. As it had already noted, in
the physics of continuous media (material systems) the interest is expressed
in generalized solutions of equations of the balance conservation laws.
These are solutions that describe the formations in material media 
observed. The investigation of relevant physical structures is carried 
out using the field-theory equations.

\bigskip
The unique properties of nonidentical evolutionary relation, which
describes the connection between physical fields and material
systems, discloses the connection of evolutionary relation with
the field-theory equations. In fact, all equations of existing
field theories are the analog to such relation or its differential
or tensor representation.

\subsection*{3. Specific features of field-theory equations}

The field-theory equations are equations that describe physical
fields. Since physical fields are formatted by physical structures,
which are described by closed exterior {\it inexact } forms and
by closed dual forms (metric forms of manifold), is obvious that
the field-theory equations or solutions to these equations have to
be connected with closed exterior forms. Nonidentical relations
for functionals like wave-function, action functional, entropy,
and others, which are obtained from the equations for material
media (and from which identical relations with closed forms
describing physical fields are obtained), just disclose the
specific features  of the field-theory equations.

The equations of mechanics, as well as the equations of continuous
media physics, are partial differential equations for desired
functions like a velocity of particles (elements), temperature,
pressure and density, which correspond to physical quantities of
material systems (continuous media). Such functions describe the
character of varying physical quantities of material system. The
functionals (and state-functions) like wave-function, action
functional, entropy and others, which specify the state of
material systems, and corresponding relations are used in
mechanics and continuous media physics only for analysis of
integrability of these equations. And in field theories such
relations play a role of equations. Here it reveals the duality of
these relations. In mechanics and continuous media physics these
equations describe the state of material systems, whereas in
field-theory they describe physical structures from which physical
fields are formatted.

{\footnotesize \{In differential equations of mathematical
physics, which describe physical processes, the functions required
are found by integrating derivatives obtained from the
differential equation. And in field-theory equations the functions
required follow not from derivatives, but from differentials
of identical relations and they are exterior forms. That is, in
mathematical physics one has to distinguish two types of
differential equations, namely, the differential equations, which
describe the variations of physical quantities, and the
field-theory equations, which describe physical structures.\}}

\bigskip

It can be shown that all equations of existing field theories are
in essence relations that connect skew-symmetric forms or their analogs
(differential or tensor ones). And yet
the nonidentical relations are treated as equations from which it can be
found identical relation with include closed forms
describing physical structures desired.

Field equations (the equations of the Hamilton formalism) reduce
to identical relation with exterior form of first degree, namely, 
to the Poincare invariant $$ds\,=-\,H\,dt\,+\,p_j\,dq_j\eqno(14)$$

{\footnotesize \{The field equation has the form [2]
$${{\partial s}\over {\partial t}}+H \left(t,\,q_j,\,{{\partial s}\over {\partial q_j}}
\right )\,=\,0,\quad
{{\partial s}\over {\partial q_j}}\,=\,p_j \eqno(15)$$
here $s$ is a field function for the action functional $S\,=\,\int\,L\,dt$.
Here $L$ is the Lagrangian function, $H(t,\,q_j,\,p_j)\,=\,p_j\,\dot q_j-L$
is the Hamilton function $p_j\,=\,\partial L/\partial \dot q_j$. These functions
satisfy the relations:
$${{dg_j}\over {dt}}\,=\,{{\partial H}\over {\partial p_j}}, \quad
{{dp_j}\over {dt}}\,=\,-{{\partial H}\over {\partial g_j}}\eqno(16)$$
Relations (16), which present a set of the Hamilton equations, are the closure
conditions for exterior and dual forms [7]. They are similar to relations
(5).\}}

The Schr\H{o}dinger equation in quantum mechanics is an analog to
field equation, where the conjugated coordinates are replaced by
operators. The Heisenberg equation corresponds to the closure
condition of dual form of zero degree. Dirac's {\it bra-} and {\it cket}- vectors made
up a closed exterior form of zero degree.
It is evident that the relations with skew-symmetric differential forms
of zero degree correspond to quantum mechanics.
The properties of skew-symmetric differential forms
of the second degree lie at the basis of the electromagnetic field
equations. The Maxwell equations may be written as $d\theta^2=0$, $d^*\theta^2=0$,
where $\theta^2= \frac{1}{2}F_{\mu\nu}dx^\mu dx^\nu$ (here
$F_{\mu\nu}$ is the strength tensor). The Einstein equation is a
relation in differential forms. This equation relates the
differential of dual form of first degree (Einstein's tensor) and
a closed form of second degree --the energy-momentum tensor. (It
can be noted that, even Einstein's equation connects the closed
forms of second degree, this equation is obtained from
differential forms of third degree).

The connection the field theory equations with skew-symmetric
forms of appropriate degrees shows that there exists a commonness
between field theories describing physical fields of different
types. This can serve as an approach to constructing the unified
field theory. This connection shows that it is possible to
introduce a classification of physical fields according to the
degree of skew-symmetric differential forms. From relations  (10)
and (13) one can see that relevant degree of skew-symmetric
differential forms, which can serve as a parameter of unified
field theory, is connected with the degree $p$ of evolutionary
form in relation (10). It should be noted that the degree $p$ is
connected with the number of interacting balance conservation
laws. {\footnotesize \{The degree of closed forms also reflects a
type of interaction [6]. Zero degree is assigned to a strong
interaction, the first one does to a weak interaction, the second
one does to electromagnetic interactions, and the third degree is
assigned to gravitational field.\}}

The connection of field-theory equations, which describe physical fields,
with the equations for material media discloses the foundations of the
general field theory. As an equation of general field theory it can serve
the evolutionary relation (10), which is obtained the balance conservation
laws for material media and has a double meaning. On the one hand, that,
being a relation, specifies the type of solutions to equations of balance conservation
laws and describes the state of material system (since it includes
the state differential), and, from other hand, that can play a role
of equations for description of physical fields (for finding the closed
inexact forms, which describe the physical structures from which physical
fields are made up). It is just a double meaning that discloses the
connection of physical fields with material media (which is based on the
conservation laws) and allows to understand on what the general field theory
has to be based.

\bigskip

In conclusion it should be emphasized that the study of equations
of mathematical physics appears to be possible due to unique
properties of skew-symmetric differential forms. In this case,
beside the exterior skew-symmetric differential forms, which 
are defined on differentiable manifolds, the skew-symmetric differential 
forms, which, unlike to the exterior forms, are defined on deforming 
(nondifferentiable) manifolds [7], were used.

1. Cartan E., Les Systemes Differentials Exterieus ef Leurs Application 
Geometriques. -Paris, Hermann, 1945.  

2. Smirnov V.~I., A course of higher mathematics. -Moscow, 
Tech.~Theor.~Lit. 1957, V.~4 (in Russian).

3. Tolman R.~C., Relativity, Thermodynamics, and Cosmology. Clarendon Press, 
Oxford,  UK, 1969.

4. Fock V.~A., Theory of space, time, and gravitation. -Moscow, 
Tech.~Theor.~Lit., 1955 (in Russian).
      
5. Clark J.~F., Machesney ~M., The Dynamics of Real Gases. Butterworths, 
London, 1964. 
 
6. Petrova L.~I. The quantum character of physical fields. Foundations 
of field theories, Electronic Journal of Theoretical Physics, v.3, 10 
(2006), 89-107p.

7. Petrova L.~I. Skew-symmetric differential forms: Conservation laws. 
Foundations of field theories. -Moscow, URSS, 2006, 158 p. (in Russian). 

\end{document}